\documentclass[]{emulateapj}
\usepackage{graphicx}
\usepackage{latexsym}
\usepackage{natbib}
\usepackage{url}

\begin{document}

\title{Discovery of an accreting millisecond pulsar in the eclipsing binary system Swift~J1749.4--2807}

\author{D. Altamirano\altaffilmark{1},
Y. Cavecchi\altaffilmark{1,2},
A. Patruno\altaffilmark{1},
A. Watts\altaffilmark{1},
M. Linares\altaffilmark{3},   
N. Degenaar\altaffilmark{1},
M. Kalamkar\altaffilmark{1},
M. van der Klis\altaffilmark{1},
N. Rea\altaffilmark{4},\
P. Casella\altaffilmark{5},
M. Armas Padilla\altaffilmark{1},
R. Kaur\altaffilmark{1},
Y.J. Yang\altaffilmark{1},
P. Soleri\altaffilmark{6},
R. Wijnands\altaffilmark{1}
}

\altaffiltext{1}{Email: d.altamirano@uva.nl ; Astronomical Institute,
  ``Anton Pannekoek'', University of Amsterdam, Science Park 904,
  1098XH, Amsterdam, The Netherlands.}  

\altaffiltext{2}{Sterrewacht Leiden, University of Leiden. (Huygens
  Laboratory J.H. Oort Building) Niels Bohrweg 2 NL-2333 CA Leiden The
  Netherlands}

\altaffiltext{3}{Massachusetts Institute of Technology - Kavli
Institute for Astrophysics and Space Research, Cambridge, MA 02139,
USA}

\altaffiltext{4}{Institut de Ciencies de l'Espai (ICE, CSIC--IEEC),
Campus UAB, Facultat de Ciencies, Torre C5-parell, 2a planta, 08193,
Bellaterra (Barcelona), Spain}

\altaffiltext{5}{School of Physics and Astronomy, University of Southampton, Southampton, Hampshire, SO17 1BJ, United Kingdom} 

\altaffiltext{6}{Kapteyn Astronomical Institute, University of
  Groningen, P.O. Box 800, 9700 AV Groningen, The Netherlands.}

\begin{abstract}

  We report on the discovery and the timing analysis of the first
  eclipsing accretion-powered millisecond X-ray pulsar (AMXP):
  SWIFT~J1749.4--2807.
  The neutron star rotates at a frequency of $\sim$517.9~Hz and is
  in a binary system with an orbital period of 8.8 hrs and a projected
  semi-major axis of $\sim$1.90 lt-s.
  Assuming a neutron star between 0.8 and 2.2 $M_{\odot}$ and using
  the mass function of the system and the eclipse half-angle, we
  constrain the mass of the companion and the inclination of the
  system to be in the $\sim$0.46-0.81 M$_{\odot}$ and 
  $\sim74.4^\circ-77.3^\circ$ range, respectively.
  To date, this is the tightest constraint on the orbital inclination
  of any AMXP.
  As in other AMXPs, the pulse profile shows harmonic content up to
  the 3rd overtone.  However, this is the first AMXP to show a 1st
  overtone with rms amplitudes between $\sim6$\% and $\sim23$\%, which
  is the strongest ever seen, and which can be more than two times
  stronger than the fundamental.
  The fact that SWIFT~J1749.4--2807 is an eclipsing system which shows
  uncommonly strong harmonic content suggests that it might be the best
  source to date to set constraints on neutron star properties including
  compactness and geometry.

\end{abstract}
 \maketitle

\section{Introduction}
\label{sec:intro}

The first accreting millisecond X-ray pulsar (hereafter AMXP) was
discovered in 1998 \citep[SAX~J1808.4--3658, see][]{Wijnands98} and
since then, a total of 13 AMXPs have been found and studied in detail
\citep{Patruno10a}.  Most AMXPs show near sinusoidal profiles during
most of their outbursts.  This is consistent with a picture in which
only one of the hotspots (at the magnetic poles) is visible (see
ref. below).  Deviations from a sinusoidal profile (i.e, an increase
in harmonic content) are generally interpreted as being caused by the
antipodal spot becoming visible, perhaps as accretion rate falls and
the disk retreats \citep[see, e.g.][ and references
therein]{Poutanen03,Ibragimov09}.

Although the amplitude of the 1st overtone may reach that of the
fundamental late in the outburst \citep[see, e.g][]{Hartman08,
  Hartman09a}, no AMXP so far has shown pulse profiles where the 1st
overtone is generally stronger than the fundamental throughout the
outburst.

The stability of the pulse profiles in some of the AMXPs means that
pulse profile modeling can be used to set bounds on the compactness of
the neutron star and hence the dense matter equation of state
\citep[see, e.g.][ and references therein]{Poutanen03,Poutanen09}.
Unfortunately, there is often a large degeneracy between the
parameters due to the number of free parameters needed to construct
the model profile. One of these parameters is the inclination of the
system, which to date has not been well-constrained for any AMXP.

\begin{figure}[http] 
\center
\resizebox{1\columnwidth}{!}{\rotatebox{0}{\includegraphics{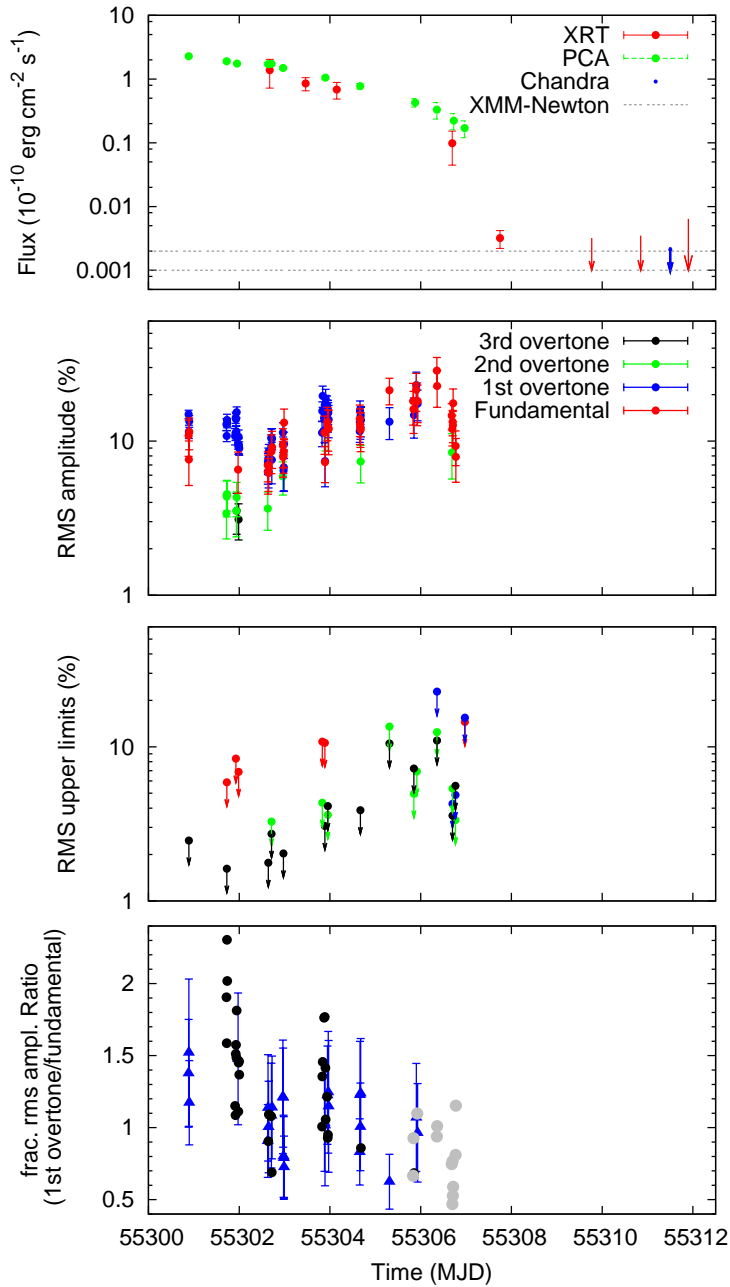}}}
{\footnotesize
\caption{\textit{Top panel:} 2-10 keV flux as measured from RXTE/PCA
  and Swift/XRT observations. The flux of the last PCA observation (MJD
  55307.5) is not shown; the spectrum of this observation was used as
  an estimate background emission (see text).
  Upper limits are quoted at 95\% confidence level.  We calculated the
  flux during the last PCA and last Swift/XRT detection using WebPIMMS
  (assuming a power law spectrum with index 1.8).
  \textit{Middle panels:} Fractional rms amplitude and 95\% confidence
  level upper limits of the fundamental and three overtones as a
  function of time. Detections ($>3\sigma$ single trial) and upper
  limits are from $\sim500$ and $\sim3000$ sec datasets, respectively.
  \textit{Bottom panel:} Ratio between the fractional rms amplitude of
  the 1st overtone and fundamental. Blue triangles
  represent points in which both harmonics are significantly detected
  in $\approx500$ second datasets.  Black circles represent the ratio
  between the fractional rms amplitude of the 1st overtone and the
  95\% confidence level upper limit to the amplitude of the
  fundamental. Grey circles represent the same ratio but when the
  fundamental is significantly detected and not the 1st overtone.
  This means that black circles represent lower limits while grey
  circles are upper limits. These ratios are independent of the
  background.}}
\label{fig:lc}
\end{figure}

In this Letter we report on the discovery and timing of the
accretion-powered millisecond X-ray pulsar SWIFT~J1749.4--2807. Thanks
to the observed eclipses \citep{Markwardt10}, we set the tightest
constraint on system inclination for any AMXP.  This, coupled with the
fact that the amplitude of the first overtone is higher/comparable to
that of the fundamental for much of the outburst and that the
amplitude of the first overtone is unusually high, allows to put tight
constraints on pulse profile models.  We show that SWIFT~J1749.4--2807
has the potential to be one of the best sources for this approach to
constraining the neutron star mass-radius relation and hence the EoS
of dense matter.
\section{SWIFT~J1749.4--2807} 

SWIFT~J1749.4--2807 was discovered in June 2, 2006 \citep{Schady06},
when a bright burst was detected by the Swift burst alert telescope
(BAT). 
\citet{Wijnands09} presented a detailed analysis of the Swift/BAT and
Swift/XRT data and showed that the spectrum of the 2006 burst was
consistent with that of a thermonuclear Type I X-ray burst \citep[see,
also]{Palmer06,Beardmore06} from a source at a distance of
$6.7\pm1.3$~kpc.

SWIFT~J1749.4--2807 was detected again between April 10th and 13th,
2010 using INTEGRAL and Swift observations
\citep{Pavan10,Chenevez10}. 
We promptly triggered approved RXTE observations on this source to
study X-ray bursts and to search for millisecond pulsations (Proposal
ID:93085-09, PI: Wijnands). The first RXTE observation was performed
on April 14th and lasted for about 1.6 ksec. We found strong coherent
pulsations at $\approx517.9$~Hz and at its first overtone
$\approx1035.8$~Hz, showing that SWIFT~J1749.4--2807 is an accreting
millisecond X-ray pulsar \citep{Altamirano10e}. RXTE followed up the
decay of the outburst on a daily basis.
Preliminary results on the rms amplitude of the pulsations, orbital
solution, discovery of eclipses, evolution of the outburst and upper
limits on the quiescent luminosity were reported in Astronomer's
Telegrams \citep{Altamirano10e, Bozzo10a, Belloni10, Strohmayer10,
  Markwardt10, Yang10, Chakrabarty10}. No optical counterpart has been
identified as yet, with a 3$\sigma$ lower limit in the i-band of
$19.6$ \citep{Yang10}.

\section{Observations, spectral analysis and background estimation}
\label{sec:dataanalysis}

We used data from the Rossi X-ray Timing Explorer (RXTE) Proportional
Counter Array \citep[PCA, for instrument information
see][]{Jahoda06}. Between April 14th and April 21st there were 15
pointed observations of SWIFT~J1749.4--2807, each covering 1 or 2
consecutive 90-min satellite orbits.

We also analyzed data from Swift's X-ray telescope
\citep[XRT;][]{Burrows05}. There were a total of 10 observations
(target ID 31686), all obtained in the Photon Counting (PC) mode.

We used standard tools and procedures to extract energy spectra from
PCA Standard 2 data. We calculated response matrices and ancillary
files for each observation using the {\tt FTOOLS} routine {\tt PCARSP
V10.1.} Background spectra were estimated using the faint-model in {\tt
  PCABACKEST} (version 6.0).
For the XRT, we used standard procedures to process and analyze the PC
mode data\footnote{http://www.swift.ac.uk/XRT.shtml}. When necessary,
an annular extraction region was used to correct for pile-up
effects. 
We generated exposure maps with the task {\tt XRTEXPOMAP} and ancillary
response files were created with {\tt XRTMKARF}. The latest response matrix
files (v. 11) were obtained from the {\tt CALDB} database.

We used an absorbed power-law to fit all PCA/XRT observations. We
first fitted all XRT spectra and found an average interstellar
absorption of $3.5\times 10^{22}$ cm$^{-2}$; $N_H$ was fixed to this
value when fitting all (standard) background-subtracted PCA spectra.
When comparing the fluxes estimated by PCA and XRT we found that
the PCA fluxes were systematically higher.
Only on one occasion RXTE and Swift observations were performed
simultaneously (MJD 55306.69, i.e. at the end of the outburst) and in
this case the flux difference was $\approx1.4 \times 10^{-10}$ erg
cm$^{-2}$ s$^{-1}$. 
This is consistent with that seen a day later.
Since (i) the count rates during the last RXTE observation (when the
source was below the PCA detection limit but detected by Swift/XRT)
are consistent with those we measure during the eclipses (see
Section~\ref{sec:eclipses}) and (ii) these count rates are consistent
with the offset we find between PCA and XRT, we conclude that there is
an additional source of background flux in our PCA observations.
To correct for this, we also use the background-corrected spectrum of
the last PCA observation (OBSID: 95085-09-02-08, MJD 55307.5) as an
estimate of the additional source of background flux.
This approach is optimal in crowded fields near the Galactic plane,
where the contribution from the Galactic ridge emission and other
X-ray sources in the 1$^\circ$ PCA FoV becomes important \citep[see,
e.g.,][]{Linares07,Linares08}.

\subsection{Background estimates and the fractional rms
  amplitudes}\label{sec:bkg}

Given the low flux during our observations, it is very important to
accurately estimate the background emission before calculating the
pulse fractional rms amplitudes.
The fact that the extra source of background photons is unknown
complicates the estimation of total background flux as a function of
time.
For example, the background flux could be intrinsically varying; even
in the case of a constant distribution of background flux in the sky,
it would be possible to measure flux variations if the collimator
(i.e. PCA) orientation on the sky changes between observations.
Given the background uncertainties, we arbitrarily adopt as total
3--16 keV background per observation the modeled background plus a
constant offset of $\approx17.5\pm2$ ct/s/PCU. This takes into account
the $\approx19.5$ ct/s/PCU as estimated by the eclipses, the last PCA
observation and the PCA$-$XRT offset (which is equivalent to
$\approx18-19$ ct/s/PCU in the 3--16 keV band as estimated with
WebPIMMS\footnote{\url{http://heasarc.gsfc.nasa.gov/Tools/w3pimms.html}}
and the best fit model to the XRT data), and the $\approx15.5$ offset
we would obtain if the additional source of background photons could
change by $\sim20$\%.
This conservative adopted possible background range results in
conservative errors on the pulsed fractions we report, i.e. the errors
are probably overestimated. 

\begin{table}
  \caption{Timing parameters for the AMXP SWIFT~J1749.4--2807 }

\scriptsize
\begin{tabular}{lc}
\hline
\hline
Parameter & Value \\
\hline
Orbital period, P$_{orb}$(days) \dotfill                           & 0.3673696(2)  \\
Projected semi major axis, $a_x \sin i$ (lt-s)\dotfill   & 1.89953(2)   \\
Time of ascending node, $T_{asc}$ (MJD) \dotfill          & 55300.6522536(7) \\
Eccentricity, e (95\% c.l.)\dotfill                                         & $<5.2 \times 10^{-5}$ \\
Spin frequency  $\nu_0$ (Hz) \dotfill                             & 517.92001395(1) \\
Pulsar mass function, $f_x$ ($M_{\sun}$)\dotfill  & $0.0545278(13)$  \\
Minimum companion mass$^1$, $M_c$ ($M_{\sun}$)\dotfill               & $0.5898$ \\
\hline
\end{tabular}\\
All errors are at $\Delta\chi^2=1$. \\
$^1$: The companion mass is estimated assuming a neutron star of 1.4 M$_{\odot}$.
\label{table:data}
\end{table}

%
%
%
%
%
%
%

\section{Outburst Evolution}

In Figure~\ref{fig:lc} we show the 2.0--10.0~keV unabsorbed flux of
SWIFT~J1749.4--2807 as measured from all available RXTE/PCA and
Swift/XRT observations.
Our dataset samples the last 7 days of the outburst, during which the
flux decayed exponentially. We find that between MJD 55306.5 and
55307.5 SWIFT~J1749.4--2807 underwent a sudden drop in flux of more
than an order of magnitude, less abrupt than the 3 orders of magnitude
drop in flux observed in the previous outburst of SWIFT~J1749.4--2807
\citep{Wijnands09}.
Similar drops in flux have been seen for other AMXPs \citep[see,
e.g.][]{Wijnands03, Patruno09d}.
If we take into account the fact that SWIFT~J1749.4--2807 was first
detected on MJD 55296 \citep{Pavan10}, we estimate an outburst
duration of about 12 days.

\begin{figure}[!hbtp] 
\resizebox{1\columnwidth}{!}{\rotatebox{-90}{\includegraphics{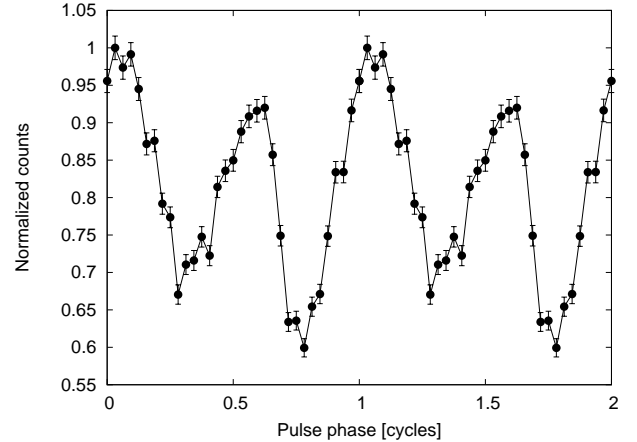}}}
\caption{Pulse profile obtained by folding $\approx\,3400$
  sec of data (ObsId 95085-09-01-02, 2-16 keV range). }
\label{fig:profile}
\end{figure}

\begin{figure}[!hbtp] 
\resizebox{1\columnwidth}{!}{\rotatebox{0}{\includegraphics{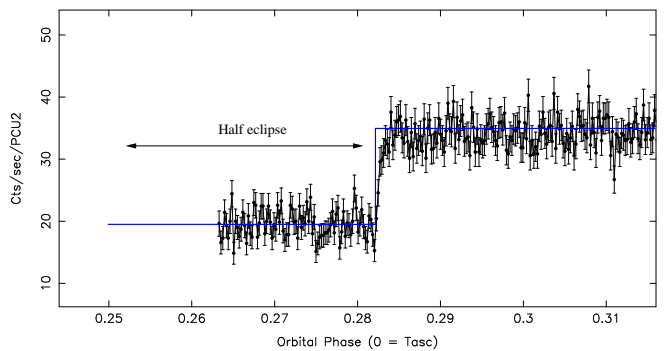}}}
\caption{Eclipse observed on observation 95085-09-02-02. The dataset
  starts on MJD 55302.9531, during the eclipse and shows that the
  egress occurs at orbital phase $\approx0.282$ (where orbital phase
  zero is the time of passage through the ascending node). We estimate
  the duration of the eclipse by assuming that eclipse is symmetric
  around orbital phase 0.25. The light curve was only corrected by the
  standard modeled background (see text). Count rates are in the 2-16
  keV band.}
\label{fig:eclipse}
\end{figure}

\section{Pulsations}

Adopting a source position $\alpha=17^h 49^m 31^s.94$, $\delta =
-28^\circ 08^{'} 05^{''}.8$ \citep[from XMM-Newton images,
see][]{Wijnands09}, we converted the photon arrival times to the Solar
System barycenter (Barycentric Dynamical Time) with the FTOOL faxbary,
which uses the JPL DE-405 ephemeris along with the spacecraft
ephemeris and fine clock corrections to provide an absolute timing
accuracy of ~3.4$\rm\,\mu s$ \citep{Jahoda06}.

We created power spectra of segments of 512 sec of data and found
strong signals at frequencies of $\approx517.92$~Hz and
$\approx1035.84$~Hz \citep{Altamirano10e}; these signals were not
always detected simultaneously with a significance greater than
$3\sigma$.

To proceed further, we used the preliminary orbital solution reported
by \citet{Strohmayer10} and folded our dataset into 87 pulse profiles
of $\approx\,500$ sec each. We then fitted the profiles with a
constant plus 4 sinusoids representing the pulse frequency and its
overtones.
We then phase-connected the pulse phases by fitting a constant pulse
frequency plus a circular Keplerian orbital model.  The procedure is
described in detail in \citet{Patruno09b}.
In Table~\ref{table:data} we report the best fit solution and
in Figure~\ref{fig:profile} we show one example of the pulse profile.

It is known that the timing residuals represent a significant
contribution to the X-ray timing noise, which if not properly taken
into account can affect the determination of the pulse frequency and
the orbital solution \citep[see, e.g. ][]{Hartman08,Patruno09b}.
There is a hint of a correlation between the X-ray timing noise and
the X-ray flux, especially between MJD 55302 and 55303, where a slight
increase of the X-ray flux is accompanied by a jump in the pulse
phases, similarly to what was reported for 6 other AMXPs by
\citet{Patruno09c}. A complete discussion of timing noise in this
source is beyond the scope of this paper and will be presented
elsewhere.

In the middle panels of Figure~\ref{fig:lc} we show the fractional rms
amplitude of the fundamental, and of the first, second and third
overtone when the signal was $>3\sigma$ significant in $\sim500$ sec
datasets. The 95\% confidence level upper limits are are estimated
using $\sim3000$ sec datasets (excluding detections) and plotted
separately for clarity.
When detected significantly, the rms amplitudes of the fundamental and
1st overtone are in the $\simeq6-29$\% and $\simeq6-23$\% ranges,
respectively; the highest values are reached at the end of the
outburst, where the uncertainties in our measurements also
increase. Amplitudes for the fundamental as high as 15-20\% rms have
been seen before for at least one source \citep[although for a brief
interval, see][]{Patruno09b}, however, no other AMXP shows a 1st
overtone as strong as we detect it in SWIFT~J1749.4--2807.
In order to compare the strength of both signals, in
Figure~\ref{fig:lc} (lower panel) we show the ratio between the
fractional rms amplitude of the 1st overtone and that of the
fundamental. As can be seen, there are periods in which the ratio is
approximately one, but also periods where the ratio is 2 or more. We
note that these ratios are independent of the uncertainties on the
background.

\section{Eclipses and the inclination of the
  system}\label{sec:eclipses}

We searched the RXTE data for the occurrence of X-ray bursts and found
none. Following \citet{Markwardt10}, we also searched for possible
signatures of eclipses and found two clear cases in the RXTE data
(ObsIDs: 95085-09-02-02 and 95085-09-02-04, beginning at MJD 55302.97
and 55305.87, respectively). PCA data on MJD 55306.97 (OBSID:
95089-09-02-11) samples an ingress, however, the count rate is too low
to extract useful information \citep[but see
][]{Markwardt10a,Ferrigno11}.

The first and clearest case of an eclipse is shown in
Figure~\ref{fig:eclipse}.  The average 3--16 keV count rate at the
beginning of the observation is about $\approx18.5-19.5$ cts/sec/PCU2
(only the standard modeled background has been subtracted) for the
first $\approx600$ sec. Then the countrate increases within a few
seconds to an average of $\approx36$ cts/sec/PCU2 and remains
approximately constant for the rest of the dataset. The other dataset
also shows a similar low-to-high count rate transition although at
lower intensities: the observation samples less than 275 sec of the
eclipse (at a rate of $\approx18-19$ cts/sec/PCU2); the count rate
after the egress is about $\approx22$ cts/sec/PCU2, i.e. much lower
than in the previous case.
Within the uncertainties on the unmodeled background, both egresses
occur between orbital phases of $\approx0.2823-0.2825$.
During the eclipse the count rates in these two observations are
consistent with the expected background of $\approx18-19$ cts/sec/PCU2
, implying that that SWIFT J1749.4--2807 most probably shows total
eclipses; however, given the uncertainties in the background (see
Section~\ref{sec:bkg}) and the sensitivity of the PCA, this should be
tested and better quantified with observations from instruments like
XMM-Newton, Suzaku or Chandra \citep[see also ][]{Pavan10,Ferrigno11}.

Using the best solution reported in Table~\ref{table:data}, we also
searched for pulsations in the 600 second period during which the
companion star is eclipsing the neutron star (see above). We found
none.  Upper limits are unconstrained.

\begin{figure}[!hbtp] 
  \resizebox{1\columnwidth}{!}{\rotatebox{0}{\includegraphics{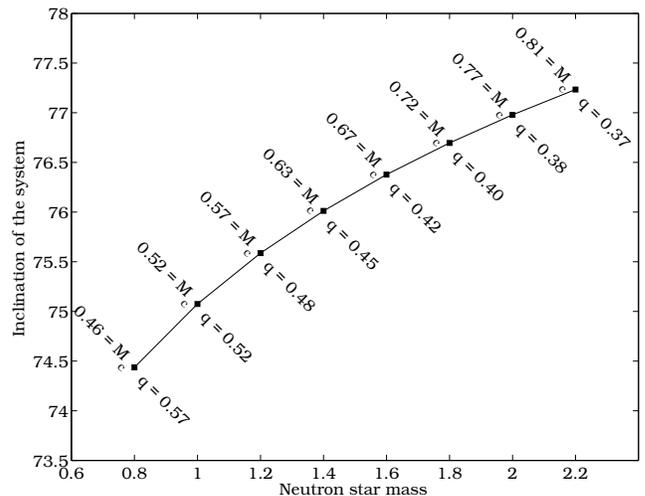}}}
  \caption{Inclination of the binary system vs. the the neutron star
    mass. For each point we also mark the mass of the companion star
    $M_c$ (in units of $M_{\odot}$) and the mass ratio
    $q=M_c/M_{NS}$.}
\label{fig:inclination}
\end{figure}

With our improved orbital solution and the measured times of the two
egresses we determine the phase of egress to be no larger than
0.2825. Assuming the eclipses are centered around neutron star
superior conjunction, the eclipse half-angle is $\approx11.7^\circ$,
corresponding to an eclipse duration of $\approx2065$ sec.

We do not detect any evidence of absorption in the form of dips in the
light curves, probably due to the fact that our dataset only samples
$\approx1.5$ orbital periods.  These dips are common in other
eclipsing LMXBs and thought to be due to the interaction of the
photons from the central X-ray emitting region by structure on the
disk rim or by what is left of the stream of incoming matter (from the
companion) above and below the accretion disk. These dips are known to
be highly energy dependent; both eclipses and egresses in our data are
energy independent.

Assuming that the companion star is a sphere with a radius R equal to
the mean Roche lobe radius, then the radius of the companion star can
be approximated as

\begin{equation}
R_L = a \cdot \frac{0.49 \cdot q^{2/3}}{0.6 \cdot q^{2/3} + \ln{(1+q^{1/3})}},
\end{equation}

where $a$ is the semi-major axis of the system and $q=M_c/M_{NS}$ is
the ratio between the companion and neutron star masses, respectively
\citep{Eggleton83}. From geometrical considerations in an eclipsing
system, if the size of the X-ray emitting region is negligible
compared with the radius of the companion star, then $R_L$ is also
related to the inclination $i$ and the eclipse half-angle $\phi$:

\begin{equation}
R_L = a \cdot \sqrt{\cos^2{i} + \sin^2{i} \cdot \sin^2{\phi}},
\end{equation}

when the eccentricity of the system is zero \citep[see, e.g.,][; also
note that the half-angle of the eclipse might be smaller as the star
filling its Roche lobe is not spherical, see,
e.g., \citealt{Chanan76}]{Chakrabarty93}.
These two equations in combination with the mass function form a
system of equations that allow us to find the inclination of the
binary system as a function of the neutron star and companion star
mass.
In Figure~\ref{fig:inclination} we show our results. For a neutron
star with mass in the 0.8-2.2 (1.4-2.2) $M_{\odot}$ range, we find
inclinations in the $74.4^{o}-77.3^{o}$ ($76.3^{o}-77.3^o$) range and
companion mass in the 0.46-0.81 (0.67-0.81) M$_{\odot}$ range.

\section{Constraining neutron star properties via pulse-profile
  modeling }

Knowing the inclination to a high degree of precision is useful for
pulse profile modeling to constrain neutron star properties including
compactness and geometry.  To explore what could be done, we tried
fitting simple model lightcurves to the pulse amplitude observations
(along the lines explored by \citealt{Pechenick83}, \citealt{Nath02}
and \citealt{Cadeau07}). The code we use has been tested against, and
is in good agreement with, the results of \citet{Lamb09a}.

We assume isotropic blackbody emission from one or two antipodal
circular hot spots, and no emission from the rest of the star or the
disc.  At this stage we ignore both Comptonization (which might be
important \citealt{Gierlinski05}) and disc obscuration.

We consider as free parameters stellar mass 
and radius, and the colatitude $\alpha$ and angular half-size $\delta$
of the hotspot(s).  Using only points where both fundamental and 1st
overtone are detected with at least $3\sigma$ significance, we search
for models that fit all observations (amplitude of fundamental and
ratio of first overtone to fundamental) and which have the same mass
and radius. Hotspot size and position are permitted to vary between
observations since accretion flow is expected to be variable.

Although it is possible to obtain a high degree of harmonic content,
due to GR effects, from a single visible hotspot (see also
\citealt{Lamb09a}), we find that the strength of the harmonic is such
that two antipodal hotspots must be visible in order to fit the data.
We are also able to constrain system geometry.  The 1$\sigma$
confidence contours restrict us to models with $\alpha \simeq
50^\circ$ and $\delta = (45-50)^\circ$; the 2$\sigma$ contours permit
a wider range of parameters but still require models where $\alpha =
(40-50)^\circ$ and $\delta = (30-50)^\circ$ (hotspots must be smaller
if they are located closer to the pole).  These results, within the
frame of our simple model, suggest a substantial offset
between rotational and magnetic pole in this source.

Our models also put limits on stellar compactness. The 1$\sigma$
confidence contours exclude models with $M/R > 0.17
M_\odot/\mathrm{km}$, while the $2\sigma$ contours exclude models with
$M/R > 0.18 M_\odot/\mathrm{km}$.  Although this does not rule out any
common equation of state \citep{Lattimer07},
it does exclude some viable regions of dense matter parameter space.

Our simple calculations, while certainly not conclusive, illustrate the
potential of this source.  With better models, and phase-resolved
spectroscopy using high spectral resolution observations, this system
is an extremely promising candidate for obtaining tight constraints
from pulse profile fitting. 

\textbf{Acknowledgments:} We thank J. Poutanen for useful
discussions. AP and ML acknowledge support from the Netherlands
Organization for Scientific Research (NWO) Veni and Rubicon
Fellowship, respectively.


\begin{thebibliography}{39}
\expandafter\ifx\csname natexlab\endcsname\relax\def\natexlab#1{#1}\fi
\expandafter\ifx\csname url\endcsname\relax
  \def\url#1{{\tt #1}}\fi
\expandafter\ifx\csname urlprefix\endcsname\relax\def\urlprefix{URL }\fi

\bibitem[{{Altamirano} et~al.(2010){Altamirano}, {Wijnands}, {van der Klis}
  et~al.}]{Altamirano10e}
{Altamirano} D., {Wijnands} R., et~al., 2010, The Astronomer's Telegram, 2565,
  1

\bibitem[{{Beardmore} et~al.(2006){Beardmore}, {Godet}, \&
  {Sakamoto}}]{Beardmore06}
{Beardmore} A.P., {Godet} O., {Sakamoto} T., 2006, GRB Coordinates Network,
  5209, 1

\bibitem[{{Belloni} et~al.(2010){Belloni}, {Stella}, {Bozzo}, {Israel}, \&
  {Campana}}]{Belloni10}
{Belloni} T., {Stella} L., et~al., 2010, The Astronomer's Telegram, 2568, 1

\bibitem[{{Bozzo} et~al.(2010){Bozzo}, {Belloni}, {Israel}, \&
  {Stella}}]{Bozzo10a}
{Bozzo} E., {Belloni} T., et~al., 2010, The Astronomer's Telegram, 2567, 1

\bibitem[{{Burrows} et~al.(2005){Burrows}, {Hill}, {Nousek} et~al.}]{Burrows05}
{Burrows} D.N., {Hill} J.E., et~al., 2005, Space Science Reviews, 120, 165

\bibitem[{{Cadeau} et~al.(2007){Cadeau}, {Morsink}, {Leahy}, \&
  {Campbell}}]{Cadeau07}
{Cadeau} C., {Morsink} S.M., et~al., 2007, \apj, 654, 458

\bibitem[{{Chakrabarty} et~al.(1993){Chakrabarty}, {Grunsfeld}, {Prince}
  et~al.}]{Chakrabarty93}
{Chakrabarty} D., {Grunsfeld} J.M., et~al., 1993, \apjl, 403, L33

\bibitem[{{Chakrabarty} et~al.(2010){Chakrabarty}, {Jonker}, \&
  {Markwardt}}]{Chakrabarty10}
{Chakrabarty} D., {Jonker} P.G., {Markwardt} C.B., 2010, The Astronomer's
  Telegram, 2585, 1

\bibitem[{{Chanan} et~al.(1976){Chanan}, {Middleditch}, \& {Nelson}}]{Chanan76}
{Chanan} G.A., {Middleditch} J., {Nelson} J.E., 1976, \apj, 208, 512

\bibitem[{{Chenevez} et~al.(2010){Chenevez}, {Brandt}, {Sanchez-Fernandez}
  et~al.}]{Chenevez10}
{Chenevez} J., {Brandt} S., et~al., 2010, The Astronomer's Telegram, 2561, 1

\bibitem[{{Eggleton}(1983)}]{Eggleton83}
{Eggleton} P.P., 1983, \apj, 268, 368

\bibitem[{{Ferrigno} et~al.(2011){Ferrigno}, {Bozzo}, {Falanga}
  et~al.}]{Ferrigno11}
{Ferrigno} C., {Bozzo} E., et~al., Jan. 2011, \aap, 525, A48+

\bibitem[{{Gierli{\'n}ski} \& {Poutanen}(2005)}]{Gierlinski05}
{Gierli{\'n}ski} M., {Poutanen} J., 2005, \mnras, 359, 1261

\bibitem[{{Hartman} et~al.(2008){Hartman}, {Patruno}, {Chakrabarty}
  et~al.}]{Hartman08}
{Hartman} J.M., {Patruno} A., et~al., 2008, \apj, 675, 1468

\bibitem[{{Hartman} et~al.(2009){Hartman}, {Patruno}, {Chakrabarty}
  et~al.}]{Hartman09a}
{Hartman} J.M., {Patruno} A., et~al., 2009, \apj, 702, 1673

\bibitem[{{Ibragimov} \& {Poutanen}(2009)}]{Ibragimov09}
{Ibragimov} A., {Poutanen} J., 2009, \mnras, 400, 492

\bibitem[{{Jahoda} et~al.(2006){Jahoda}, {Markwardt}, {Radeva}
  et~al.}]{Jahoda06}
{Jahoda} K., {Markwardt} C.B., et~al., 2006, \apjs, 163, 401

\bibitem[{{Lamb} et~al.(2009){Lamb}, {Boutloukos}, {Van Wassenhove}
  et~al.}]{Lamb09a}
{Lamb} F.K., {Boutloukos} S., et~al., 2009, \apj, 706, 417

\bibitem[{{Lattimer} \& {Prakash}(2007)}]{Lattimer07}
{Lattimer} J.M., {Prakash} M., 2007, \physrep, 442, 109

\bibitem[{{Linares} et~al.(2007){Linares}, {van der Klis}, \&
  {Wijnands}}]{Linares07}
{Linares} M., {van der Klis} M., {Wijnands} R., 2007, \apj, 660, 595

\bibitem[{{Linares} et~al.(2008){Linares}, {Wijnands}, {van der Klis}
  et~al.}]{Linares08}
{Linares} M., {Wijnands} R., et~al., 2008, \apj, 677, 515

\bibitem[{{Markwardt} \& {Strohmayer}(2010)}]{Markwardt10a}
{Markwardt} C.B., {Strohmayer} T.E., 2010, \apjl, 717, L149

\bibitem[{{Markwardt} et~al.(2010){Markwardt}, {Strohmayer}, {Swank},
  {Pereira}, \& {Smith}}]{Markwardt10}
{Markwardt} C.B., {Strohmayer} T.E., et~al., 2010, The Astronomer's Telegram,
  2576, 1

\bibitem[{{Nath} et~al.(2002){Nath}, {Strohmayer}, \& {Swank}}]{Nath02}
{Nath} N.R., {Strohmayer} T.E., {Swank} J.H., 2002, \apj, 564, 353

\bibitem[{{Palmer} et~al.(2006){Palmer}, {Barbier}, {Barthelmy}
  et~al.}]{Palmer06}
{Palmer} D., {Barbier} L., et~al., 2006, GRB Coordinates Network, 5208, 1

\bibitem[{{Patruno}(2010)}]{Patruno10a}
{Patruno} A., 2010, Proceedings of Science: "High Time Resolution Astrophysics
  IV - The Era of Extremely Large Telescopes - HTRA-IV" -- astro-ph/1007.1108

\bibitem[{{Patruno} et~al.(2009{\natexlab{a}}){Patruno}, {Watts}, {Klein Wolt},
  {Wijnands}, \& {van der Klis}}]{Patruno09d}
{Patruno} A., {Watts} A., et~al., 2009{\natexlab{a}}, \apj, 707, 1296

\bibitem[{{Patruno} et~al.(2009{\natexlab{b}}){Patruno}, {Wijnands}, \& {van
  der Klis}}]{Patruno09c}
{Patruno} A., {Wijnands} R., {van der Klis} M., 2009{\natexlab{b}}, \apjl, 698,
  L60

\bibitem[{{Patruno} et~al.(2010){Patruno}, {Hartman}, {Wijnands},
  {Chakrabarty}, \& {van der Klis}}]{Patruno09b}
{Patruno} A., {Hartman} J.M., et~al., Jul. 2010, \apj, 717, 1253

\bibitem[{{Pavan} et~al.(2010){Pavan}, {Chenevez}, {Bozzo} et~al.}]{Pavan10}
{Pavan} L., {Chenevez} J., et~al., 2010, The Astronomer's Telegram, 2548, 1

\bibitem[{{Pechenick} et~al.(1983){Pechenick}, {Ftaclas}, \&
  {Cohen}}]{Pechenick83}
{Pechenick} K.R., {Ftaclas} C., {Cohen} J.M., 1983, \apj, 274, 846

\bibitem[{{Poutanen} \& {Gierli{\'n}ski}(2003)}]{Poutanen03}
{Poutanen} J., {Gierli{\'n}ski} M., 2003, \mnras, 343, 1301

\bibitem[{{Poutanen} et~al.(2009){Poutanen}, {Ibragimov}, \&
  {Annala}}]{Poutanen09}
{Poutanen} J., {Ibragimov} A., {Annala} M., 2009, \apjl, 706, L129

\bibitem[{{Schady} et~al.(2006){Schady}, {Beardmore}, {Marshall}
  et~al.}]{Schady06}
{Schady} P., {Beardmore} A.P., et~al., 2006, GRB Coordinates Network, 5200, 1

\bibitem[{{Strohmayer} \& {Markwardt}(2010)}]{Strohmayer10}
{Strohmayer} T.E., {Markwardt} C.B., 2010, The Astronomer's Telegram, 2569, 1

\bibitem[{{Wijnands} \& {van der Klis}(1998)}]{Wijnands98}
{Wijnands} R., {van der Klis} M., 1998, \nat, 394, 344

\bibitem[{{Wijnands} et~al.(2003){Wijnands}, {van der Klis}, {Homan}
  et~al.}]{Wijnands03}
{Wijnands} R., {van der Klis} M., et~al., 2003, \nat, 424, 44

\bibitem[{{Wijnands} et~al.(2009){Wijnands}, {Rol}, {Cackett}, {Starling}, \&
  {Remillard}}]{Wijnands09}
{Wijnands} R., {Rol} E., et~al., 2009, \mnras, 393, 126

\bibitem[{{Yang} et~al.(2010){Yang}, {Russell}, {Wijnands} et~al.}]{Yang10}
{Yang} Y.J., {Russell} D.M., et~al., 2010, The Astronomer's Telegram, 2579, 1

\end{thebibliography}

\end{document}